\documentclass{PoS}

\title{Wigner Distributions and Orbital Angular Momentum of Quarks and
Gluons }

\ShortTitle{Wigner Distributions}

\author{\speaker{Asmita Mukherjee}\\
        Department of Physics, Indian Institute of Technology Bombay, Powai,
Mumbai 400076, India\\
        E-mail: \email{asmita@phy.iitb.ac.in}}

\abstract{We present a recent calculation of the Wigner distributions of quarks and
gluons in a perturbative model. We also present the results for the orbital
angular momentum and the spin-orbit correlations.}

\FullConference{QCD Evolution 2015 -QCDEV2015-\\
		26-30 mAY 2015\\
		Jefferson Lab (JLAB), Newport News Virginia, USA}

\begin{document}

\section{Introduction}
Recently the Wigner distributions of quarks and gluons inside the nucleon
\cite{ji,ji2} have been proposed as a tool to access the still unknown
orbital angular momentum of quarks and gluons and its contribution to the
nucleon spin. The 'nucleon spin problem' is till now one of the most intriguing
problems in hadron physics that has to be understood. It was found in the
EMC experiment \cite{emc} that the quarks carry about 25 \% of the nucleon spin, which
suggested that a large fraction of the contributions come from the
intrinsic spin of the quarks and gluons and their orbital angular momentum
(OAM). There have been  tremendous efforts and advances both in theoretical
and experimental sides to unravel the different components and contributions 
to the proton spin. Most recently, a global analysis  \cite{fit} including high
statistics data from STAR and PHENIX collaborations at RHIC, BNL found that
the gluon spin contribution may be about 35 \% of the proton spin. This
result has a lot of uncertainties in the small x region. In any case, quark
and gluon OAM play a substantial role in building up the proton spin.

An issue that complicates the understanding of the nucleon spin in terms of
its different components is gauge invariance. There are two main
decomposition of nucleon spin.  In the so-called canonical decomposition
\cite{jm},
one writes the nucleon helicity  in terms of the intrinsic quark spin
($\Delta q$), intrinsic gluon spin ($\Delta g$) and quark and gluon OAM
($l_{q/g}$). Except for the quark spin, it has been shown that the
other terms depend on the choice of the gauge. In the kinetic decomposition
of the nucleon spin \cite{ji1}, one writes the nucleon helicity in terms of the
intrinsic quark spin ($\Delta q$), quark OAM ($L^q$), and the total
contribution of the gluon angular momentum ($J^g$). Wakamatsu \cite{wak} separated         
$J^g$ in the kinetic decomposition into an orbital part and an intrinsic
part using a prescription similar to \cite{chen}. The kinetic and canonical OAM
differ in the choice of Wilson line needed for the color gauge invariance.
A physical interpretation of both types of OAM can be found in \cite{mb}.

Wigner distributions, which are quasiprobabilistic distributions, are known
in the context of quantum mechanics since a long time \cite{wigner}. 
Because of Heisenberg
uncertainty principle, a quantum phase space description is not possible for
observables. Wigner distributions, which are joint position and momentum
space distributions do not have probabilistic interpretation. Wigner
distributions for the quarks and gluons in the nucleon are 6 dimensional
objects, 3 positions and 3 momentum. By integrating out one or more
variables, one obtains the reduced Wigner distributions. These can have
probabilistic interpretation. Wigner distributions in the light-cone or
infinite momentum frame were introduced in \cite{lorce}. These are related to the
generalized parton correlation functions (GPCFs), which are fully
unintegrated off-forward parton correlators that contain maximum amount of
information on the correlations of quarks and gluons inside the nucleon.
Integrating the GPCFs over the light cone energy $k^-$ one gets the
generalized transverse momentum dependent pdfs (GTMDs) \cite{metz}. These GTMDs can be
expressed as Fourier transforms of the Wigner distributions. Integrating over
more variables, one can connect the Wigner distributions to the generalized
parton distribution (GPDs) and transverse momentum dependent distributions
(TMDs), both of which are known to give important information on the
momentum and angular momentum correlations of the quarks and gluons inside
the nucleon. Thus these are called 'mother distributions' containing maximum
information on the internal structure of the nucleons.             

Wigner distributions can be related to the OAM of the quarks and gluons as
well as their spin-orbit correlations. In this way these can give informations  
beyond those that can be obtained from TMDs and GPDs. As the Wigner
distributions themselves cannot be directly measured experimentally,
informations about them can be obtained indirectly through the measurement
of other observables. That is why  model calculations of Wigner distributions are
interesting as these give valuable insight into these objects.
Models in which Wigner distributions have been investigated in the
literature include constituent quark model and chiral quark soliton model
\cite{lorce}.
Both of these are phenomenological models of the nucleon and they do not
contain any gluonic degrees of freedom. We present a recent calculation of
the Wigner distributions in a perturbative model with a gluonic degree of
freedom, namely a quark dressed with a gluon \cite{ours1,ours2}. We use light-front Hamiltonian
approach and express the Wigner distributions in terms of overlaps of
light-front wave function (LFWFs). The state can be thought of as a spin 1/2
relativistic composite object. This approach gives an intuitive picture of
deep inelastic processes; it is based on field theory but at the same time
keeps close contact with the parton model ideas \cite{hari}, the field theoretic partons
have intrinsic transverse momenta and they interact. The advantage is that as
there are gluonic degrees of freedom it is possible to investigate both
quark and gluon Wigner distributions.

%%%%%%%%%%%%%%%%%%%%%%%%%%%%%%%%%%%%%%%%%%%%%%%%%%%%%%%%%%%%%%
\section{Wigner distributions for the quarks and gluons}
%%%%%%%%%%%%%%%%%%%%%%%%%%%%%%%%%%%%%%%%%%%%%%%%%%%%%%%%%%%%%
The Wigner distribution of quarks can be defined as the two-dimensional Fourier transforms
\cite{lorce,metz}

\begin{eqnarray} 
\label{main}
\rho^{[\Gamma]} ({b}_{\perp},{k}_{\perp},x,\sigma) = \int \frac{d^2
\Delta_{\perp}}
{(2\pi)^2} e^{-i \Delta_{\perp}.b_{\perp}}
W^{[\Gamma]} (\Delta_{\perp},{k}_{\perp},x,\sigma);
\end{eqnarray}
 where  $\Delta_{\perp}$ is momentum transfer of the target in transverse
direction and ${b}_{\perp}$ is the impact parameter 
conjugate to $\Delta_{\perp}$. $W^{[\Gamma]}$ is
the quark-quark correlator:
\begin{eqnarray} 
\label{qqc}
W^{[\Gamma]} ({\Delta}_{\perp},{k}_{\perp},x,\sigma)  =  \Big{\langle
}p^{+},
\frac{\Delta_{\perp}}{2},\sigma  \Big{|}
W^{[\Gamma]} (0_{\perp},k_{\perp},x)  \Big{|}
p^{+},-\frac{\Delta_{\perp}}{2},\sigma \Big{\rangle }
 \nonumber \\ \nonumber \\
=\frac{1}{2}\int\frac{dz^{-}d^{2} z_{\perp}}{(2\pi)^3}e^{i(xp^+
z^-/2-k_{\perp}.z_{\perp})}
 \Big{\langle } p^{+},\frac{\Delta_{\perp}}{2},\sigma \Big{|}
\overline{\psi}(-\frac{z}{2}) \Omega\Gamma \psi(\frac{z}{2}) \Big{|}
p^{+},-\frac{\Delta_{\perp}}{2},\sigma \Big{\rangle }
\Big{|}_{z^{+}=0}.
\end{eqnarray} \\
We use  the symmetric frame where $p^+$ and
$\sigma$ define the longitudinal momentum of the target state and its
helicity respectively. $x = k^+/p^+$ is the fraction of
longitudinal momentum fraction carried by the active quark. $\Omega$ is the
gauge link needed for color gauge invariance. In this work, we use the
light-front gauge and take the gauge link to be unity. The symbol $\Gamma$
is the Dirac matrix defining the types of quark densities; we take $\Gamma=
\gamma^+$ and $\gamma^+ \gamma^5$ respectively. 
\\%

We denote the different Wigner distributions by 
$\rho_{\lambda\lambda^\prime}$ \cite{lorce},
where $\lambda(\lambda^\prime)$ is
longitudinal polarization of target state(quark).

\begin{eqnarray} \label{ruu}
\rho_{UU}({b}_\perp,{k}_\perp,x) = \frac{1}{2}\Big[\rho^{[\gamma^+]}
({b}_\perp,{k}_\perp,x,+{e}_z) +
\rho^{[\gamma^+]}({b}_\perp,{k}_\perp,x,-{e}_z) \Big]
\end{eqnarray}
gives the Wigner distribution of unpolarized quarks in the unpolarized target
state.

\begin{eqnarray} \label{rlu}
\rho_{LU}({b}_\perp,{k}_\perp,x) = \frac{1}{2}\Big[\rho^{[\gamma^+]}
({b}_\perp,{k}_\perp,x,+{e}_z) -
\rho^{[\gamma^+]}({b}_\perp,{k}_\perp,x,-{e}_z) \Big]
\end{eqnarray}
gives the distortion due to the longitudinal polarization of  the target state.

\begin{eqnarray} \label{rul}
\rho_{UL}({b}_\perp,{k}_\perp,x) = \frac{1}{2}\Big[\rho^{[\gamma^+\gamma_5]}
({b}_\perp,{k}_\perp,x,+{e}_z)+
\rho^{[\gamma^+\gamma_5]}({b}_\perp,{k}_\perp,x,-{e}_z) \Big]
\end{eqnarray}
represents the distortion due to the longitudinal polarization of quarks.

\begin{eqnarray} \label{rll}
\rho_{LL}({b}_\perp,{k}_\perp,x) = \frac{1}{2}\Big[\rho^{[\gamma^+\gamma_5]}
({b}_\perp,{k}_\perp,x,+{e}_z)-
\rho^{[\gamma^+\gamma_5]}({b}_\perp,{k}_\perp,x,-{e}_z) \Big]
\end{eqnarray}
gives the distortion due to the correlation between the longitudinal
polarized target state and quarks. ${e_z}$ correspond to helicity of the target
state.

The Wigner distribution for the gluons can be defined as \cite{jifeng}

\begin{eqnarray}
x W^{g}(x,\vec{k}_{\perp},\vec{b}_{\perp})=\int \frac{d^2
\vec{\Delta}_{\perp}}{(2\pi)^2}
e^{-i\vec{\Delta}_{\perp}.\vec{b}_{\perp}} \int
\frac{dz^{-}d^{2} z_{\perp}}{2(2\pi)^3 p^+}e^{i k.z}
 \nonumber \\ \Big{\langle } p^{+},\frac{\vec{\Delta}_{\perp}}{2},\sigma \Big{|}
\Gamma^{ij} F^{+i}\Big( -\frac{z}{2}\Big) F^{+j}\Big( \frac{z}{2}\Big)
\Big{|}
p^{+},-\frac{\vec{\Delta}_{\perp}}{2},\sigma  \Big{\rangle }
\Big{|}_{z^{+}=0};
\label{eq1} 
\end{eqnarray}

We calculate Eq.~(\ref{eq1}) for $\Gamma^{ij} = \delta^{ij}$ ($W_1$) and
$\Gamma^{ij} = -i\epsilon^{ij}_{\perp}$ ($W_2$). We choose the light-front gauge,
and like in the quark case, take the gauge link to be unity.

We consider only longitudinally polarized target state and then we have four
gluon Wigner distributions as follows, in
a manner similar to quark Wigner distributions \cite{lorce}

\noindent
Wigner distribution of unpolarized gluon in unpolarized target state is
defined as
\begin{eqnarray}
W^{UU}=W^{\uparrow \uparrow }_{1}(x,k_{\perp},b_{\perp})+ W^{\downarrow
\downarrow }_{1}(x,k_{\perp},b_{\perp});
\end{eqnarray}
\noindent
Wigner distribution corresponding to the distortion due to longitudinal
polarization of the target:
\begin{eqnarray}
W^{LU}=W^{\uparrow \uparrow }_{1}(x,k_{\perp},b_{\perp})
-W^{\downarrow \downarrow }_{1}(x,k_{\perp},b_{\perp});
\end{eqnarray}
\noindent
Wigner distribution for  the distortion due to longitudinal
polarization of the gluons is given by :
\begin{eqnarray}
W^{UL}=W^{\uparrow \uparrow }_{2}(x,k_{\perp},b_{\perp})
+W^{\downarrow \downarrow }_{2}(x,k_{\perp},b_{\perp});
\end{eqnarray}
\noindent
and the Wigner distribution describing the correlation due to
longitudinal polarization
of the target state and the gluons
\begin{eqnarray}
W^{LL}=W^{\uparrow \uparrow }_{2}(x,k_{\perp},b_{\perp})
-W^{\downarrow \downarrow }_{2}(x,k_{\perp},b_{\perp}).
\end{eqnarray}

We present a calculation of the above Wigner distributions for  a quark state
dressed with a gluon, in a perturbative model. The quark state of momentum 
$p$ and helicity $\sigma$, can be expanded in Fock space in terms of
multi-parton light-front wave functions (LFWFs) \cite{hari}
\begin{eqnarray} 
\label{dqs}
  \Big{| }p^{+},p_{\perp},\sigma  \Big{\rangle} = \Phi^{\sigma}(p)
b^{\dagger}_{\sigma}(p)
 | 0 \rangle +
 \sum_{\sigma_1 \sigma_2} \int [dp_1]
 \int [dp_2] \sqrt{16 \pi^3 p^{+}}
 \delta^3(p-p_1-p_2) \nonumber \\ \Phi^{\sigma}_{\sigma_1 \sigma_2}(p;p_1,p_2)
b^{\dagger}_{\sigma_1}(p_1)
 a^{\dagger}_{\sigma_2}(p_2)  | 0 \rangle;
\end{eqnarray}
 
here $[dp] =  \frac{dp^{+}d^{2}p_{\perp}}{ \sqrt{16 \pi^3 p^{+}}}$. $
\Phi^{\sigma}(p)$ and
$ \Phi^{\sigma}_{\sigma_1 \sigma_2}$ are the single particle (quark) and two
particle 
(quark-gluon) LFWFs, respectively;
$\sigma_1$ and $\sigma_2$ are the helicities of the quark and gluon.
$ \Phi^{\sigma}(p)$ gives  the normalization of he state. The two particle LFWF
is related to the boost invariant LFWF; $\Psi^{\sigma}_{\sigma_1
\sigma_2}(x, q_\perp) =   
\Phi^{\sigma}_{\sigma_1 \sigma_2}
\sqrt{P^+}$.  Here we have used the relation:
\begin{eqnarray}
p_i^+= x_i p^+, ~~~~~~~~~~q_{i \perp}= k_{i \perp}+x_i p_\perp
\end{eqnarray}
and $\sum_i x_i=1, \sum_i q_{i\perp}=0$.
These two-particle LFWFs  be calculated perturbatively \cite{hari}. Using a
particular representation of the $\gamma$ matrices, the LFWFs can be written
in a two component formalism \cite{zhang}. For a bound state, the bound state mass $M$
should be less than the sum of the masses of the constituents for stability.
However, we use the same mass for the bare
as well as the dressed quark  \cite{hari}. 
As stated before, the single particle sector contributes through the normalization
of the state, which is important to get the contribution at $x=1$.
We restrict ourselves to the kinematic region $x<1$, and for our purpose 
the contribution from $ \Phi^{\sigma}(p)$ can be taken to be $1$.
\begin{figure}[!htp]   
\begin{minipage}[c]{1\textwidth}
\tiny{(a)}\includegraphics[width=6.8cm,height=6cm,clip]{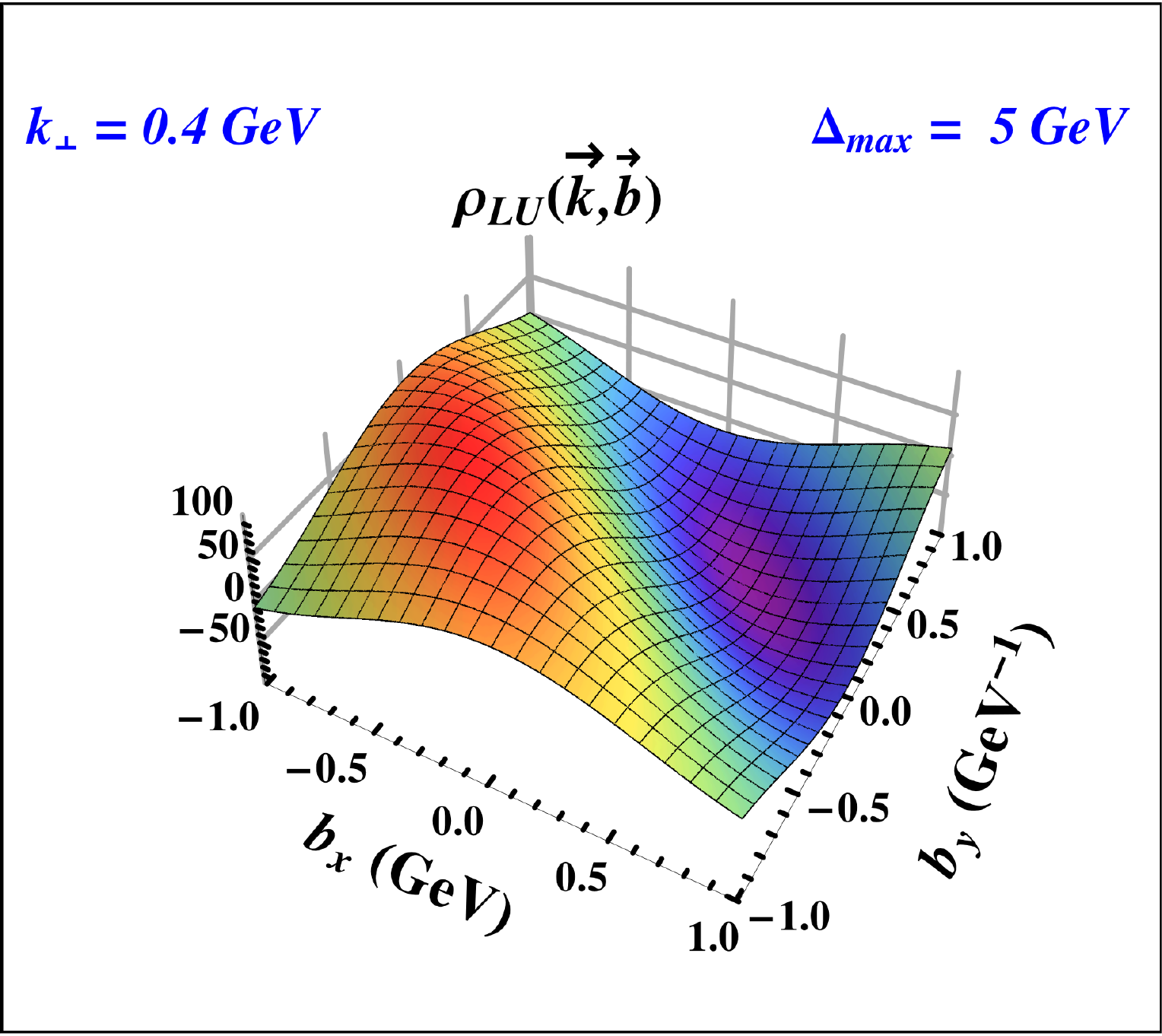}
\hspace{0.1cm}
\tiny{(b)}\includegraphics[width=6.8cm,height=6cm,clip]{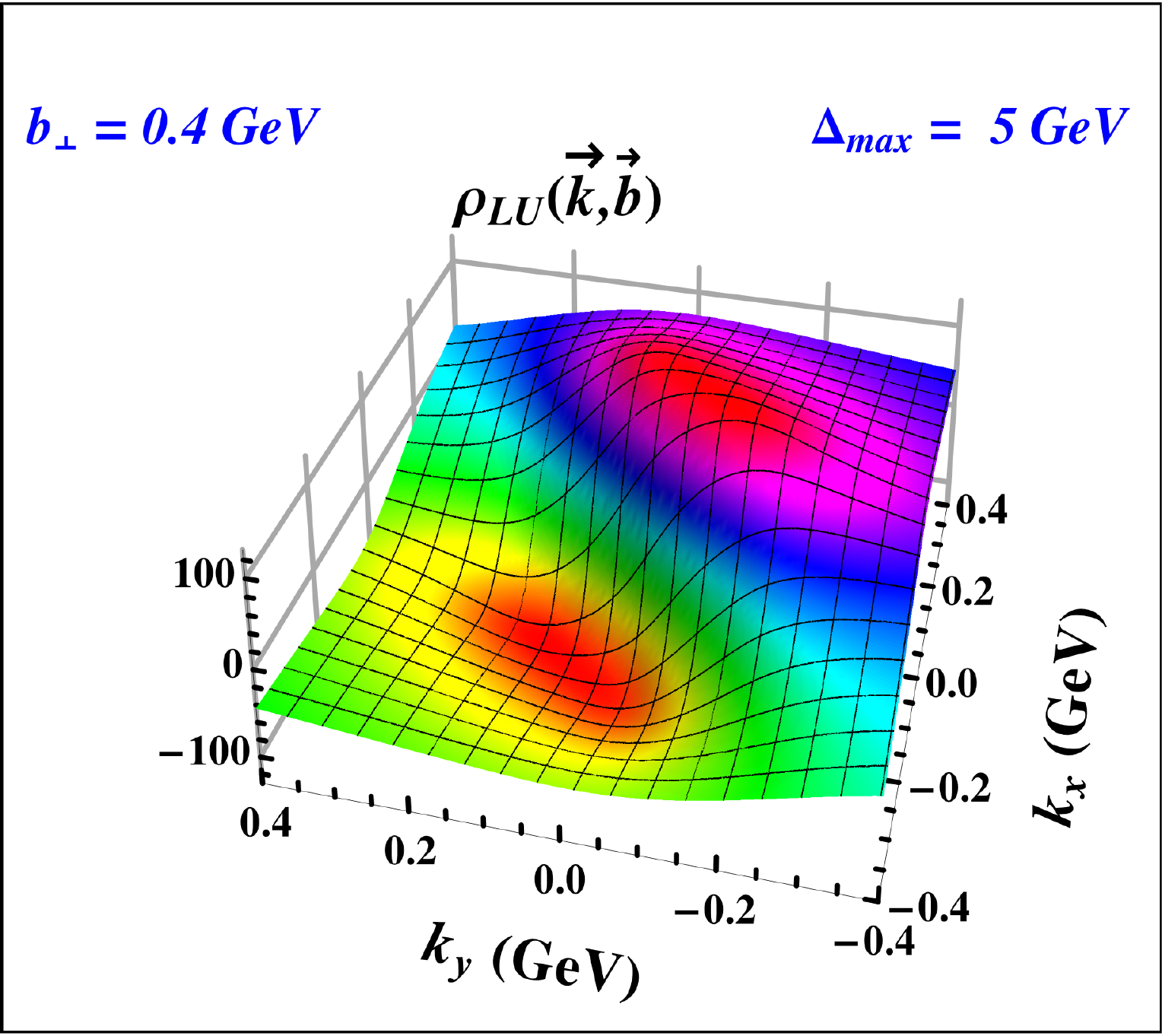}
\end{minipage}
\caption{\label{fig1}(Color online)
3D plots of the Wigner distributions $\rho_{LU}$. The plot (a) is in
$b$ space with $k_\perp = 0.4$ GeV. Plot (d) is in $k$ space with 
$b_\perp = 0.4$ $GeV^{-1}$. We took $\Delta_{max} = 5.0$ GeV.
For all the plots we kept $m = 0.33$ GeV, integrated out the $x$ variable
and we took $\vec{k_\perp} = k \hat{j}$ and $\vec{b_\perp} = b \hat{j}$
\cite{ours1}.}
\end{figure}

\begin{figure}[!htp]
\begin{minipage}[c]{1\textwidth}
\tiny{(a)}\includegraphics[width=6.8cm,height=6cm,clip]{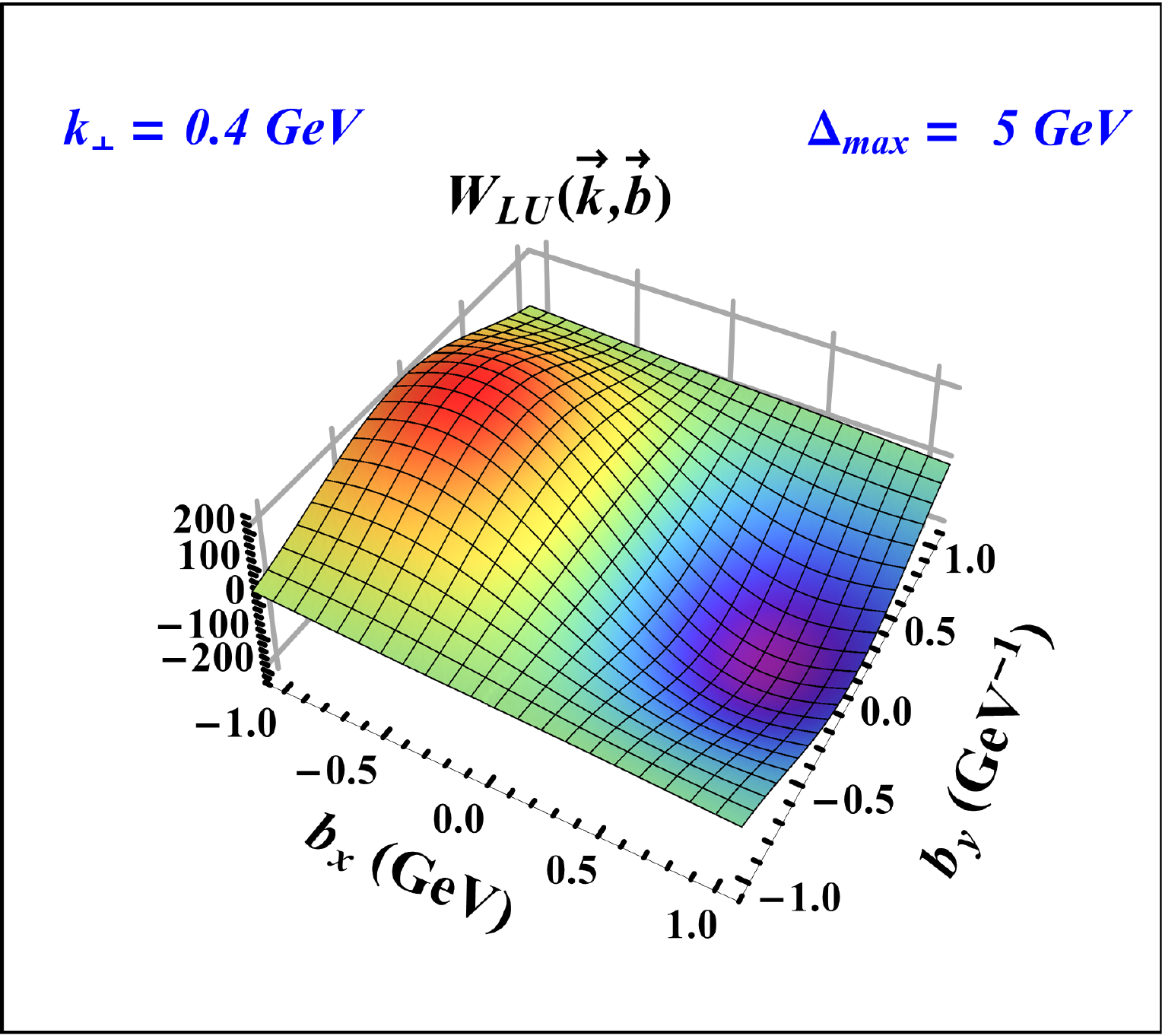}
\hspace{0.1cm}
\tiny{(b)}\includegraphics[width=6.8cm,height=6cm,clip]{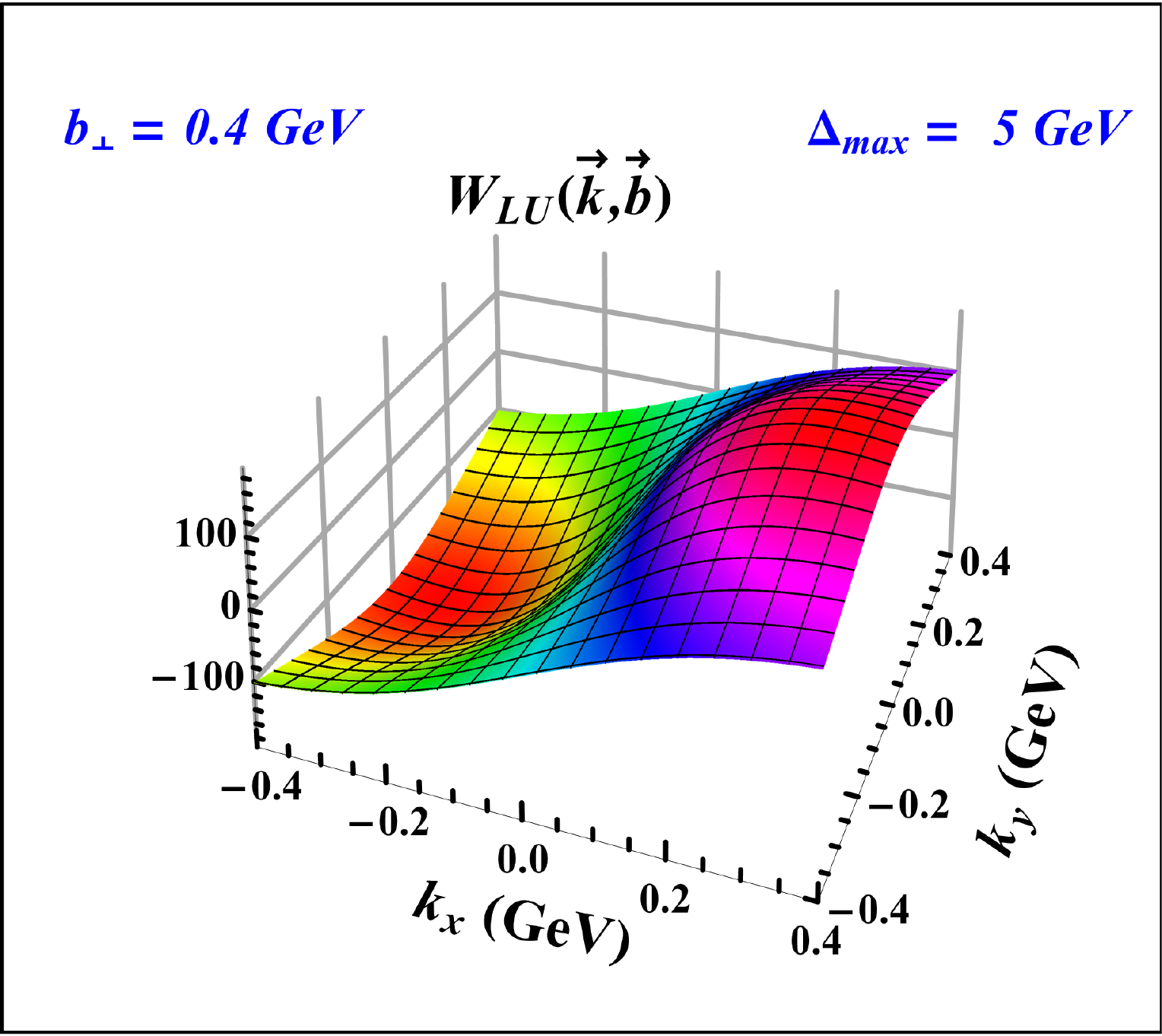}
\end{minipage}
\caption{\label{fig2}(Color online) 3D plots of the Wigner distributions $W^{LU}$. 
Plot (a) is in $b$ space with 
$k_\perp = 0.4$ GeV. Plot (b) is in $k$ space with $b_\perp = 0.4$ 
$ \mathrm{GeV}^{-1}$. We have taken $\Delta_{max} = 5.0$ GeV.
For all the plots we kept $m = 0.33$ GeV, integrated out the $x$ variable
and we took $\vec{k_\perp} = k \hat{j}$ and $\vec{b_\perp} = b \hat{j}$
\cite{ours2}.}
\end{figure}

In Figs.~\ref{fig1} and \ref{fig2}, we have shown the 3D plots for the Wigner
distributions $\rho_{LU}$ for the quarks and $W^{LU}$ for the gluons 
respectively in the impact parameter space
($b_x$-$b_y$) and in  momentum  space ($k_x$-$k_y$). The plots are taken from \cite{ours1,ours2}.
Normally the upper limit of the Fourier transform should be infinite. But in
our numerical calculation, we chose an upper limit of $\mid \Delta_\perp \mid $
called $\Delta_{max}$. The peak of the Wigner distribution increases in
magnitude as $\Delta_{max}$ increases. For all the plots, we have
taken mass of target state to be 0.33 GeV. Also we integrated over $x$ and divided by
a normalization constant. $\rho_{LU}$ is the distortion of the Wigner distribution of 
unpolarized quarks due to the longitudinal polarization of the dressed
quark.
We observe a dipole structure in $b_\perp$ space, similar to that observed
in other models. For the gluon distribution $W^{LU}$ we observe a similar
behaviour. 

%%%%%%%%%%%%%%%%%%%%%%%%%%%%%%%%%%%%%%%%%%%%%%%%%%%%%%%%%%%%%%
\section{Orbital angular momentum of quarks and gluons}
%%%%%%%%%%%%%%%%%%%%%%%%%%%%%%%%%%%%%%%%%%%%%%%%%%%%%%%%%%%%%%

The kinetic OAM for the quarks is given in terms of the GPDs \cite{ji1} as :
\begin{eqnarray}
L^{q}_{z} = \frac{1}{2} \int dx \{
x [ H^{q}(x,0,0) + E^{q}(x,0,0) ] - \tilde{H^q}(x,0,0)
\}. \nonumber
\end{eqnarray}
\\
\noindent
The GPDs in the above equation are defined with  the momentum
transfer purely in the transverse direction. Using the GPDs in our model,  we
can calculate the kinetic quark OAM. The kinetic OAM is also related to the GTMDs
\cite{metz} by the following relations:
\begin{eqnarray}
  H^q(x,0,t) = \int d^{2} k_{\perp} F_{11};  \\
 E^q(x,0,t) = \int d^{2} k_{\perp} \Big[
 -F_{11} +2\Big(
 \frac{k_{\perp}.\Delta_{\perp}}{\Delta_{\perp}^{2}} F_{12} + F_{13}
 \Big)
 \Big];  \\
 \tilde{H}^q(x,0,t) = \int d^{2} k_{\perp} G_{14}.
  \end{eqnarray} \\

\begin{figure}[!htp]
\begin{minipage}[c]{1\textwidth}
\tiny{(a)}\includegraphics[width=7cm,height=6.5cm,clip]{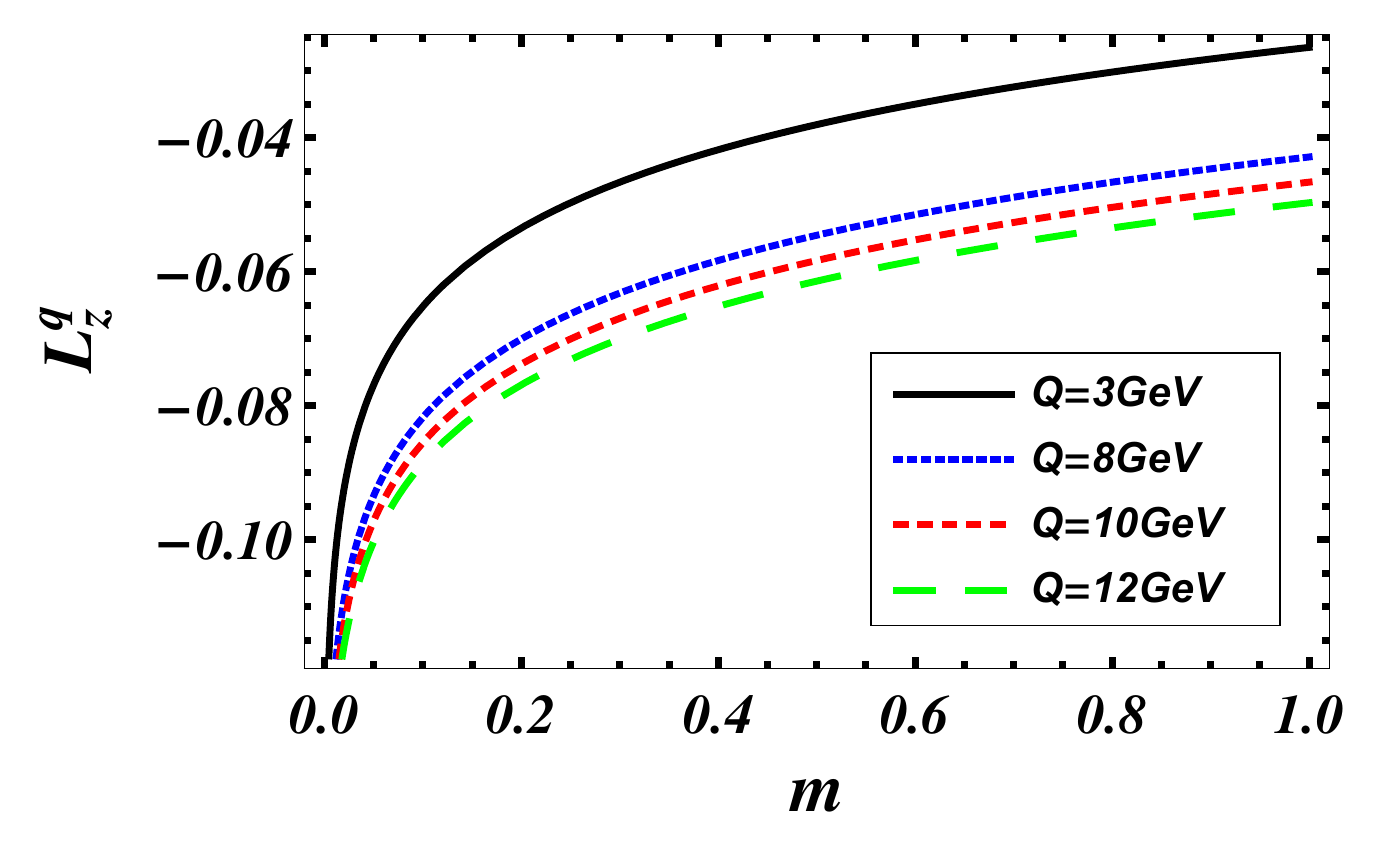}
\hspace{0.1cm}
\tiny{(b)}\includegraphics[width=7cm,height=6.5cm,clip]{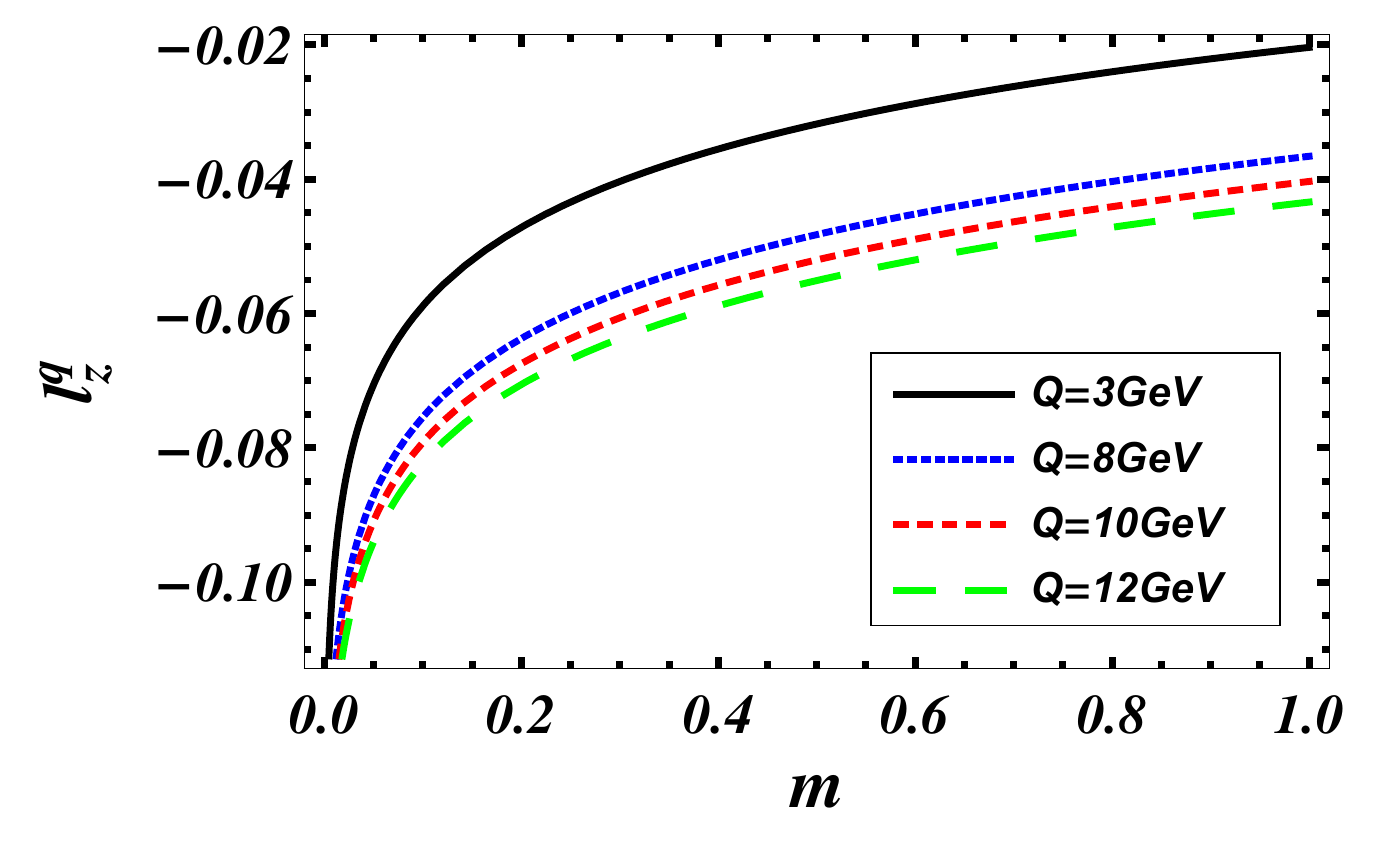}
\end{minipage}
\caption{\label{fig3}(Color online)
Plots of quark OAM (a) kinetic  and (b) canonical  vs mass of the active
quark (GeV). $Q$ is the upper limit of the transverse momentum in GeV
\cite{ours1}}
\end{figure}

The GTMD  $F_{14}$ is related to the canonical OAM as shown in
\cite{lorce, hatta1, lorce2}:

\begin{eqnarray}
l^{q}_{z} = -\int dx d^{2}k_{\perp} \frac{k_{\perp}^2}{m^2} F_{14}.
\end{eqnarray}

\begin{figure}[!htp]
\begin{minipage}[c]{1\textwidth}
\tiny{(a)}\includegraphics[width=7cm,height=6.5cm,clip]{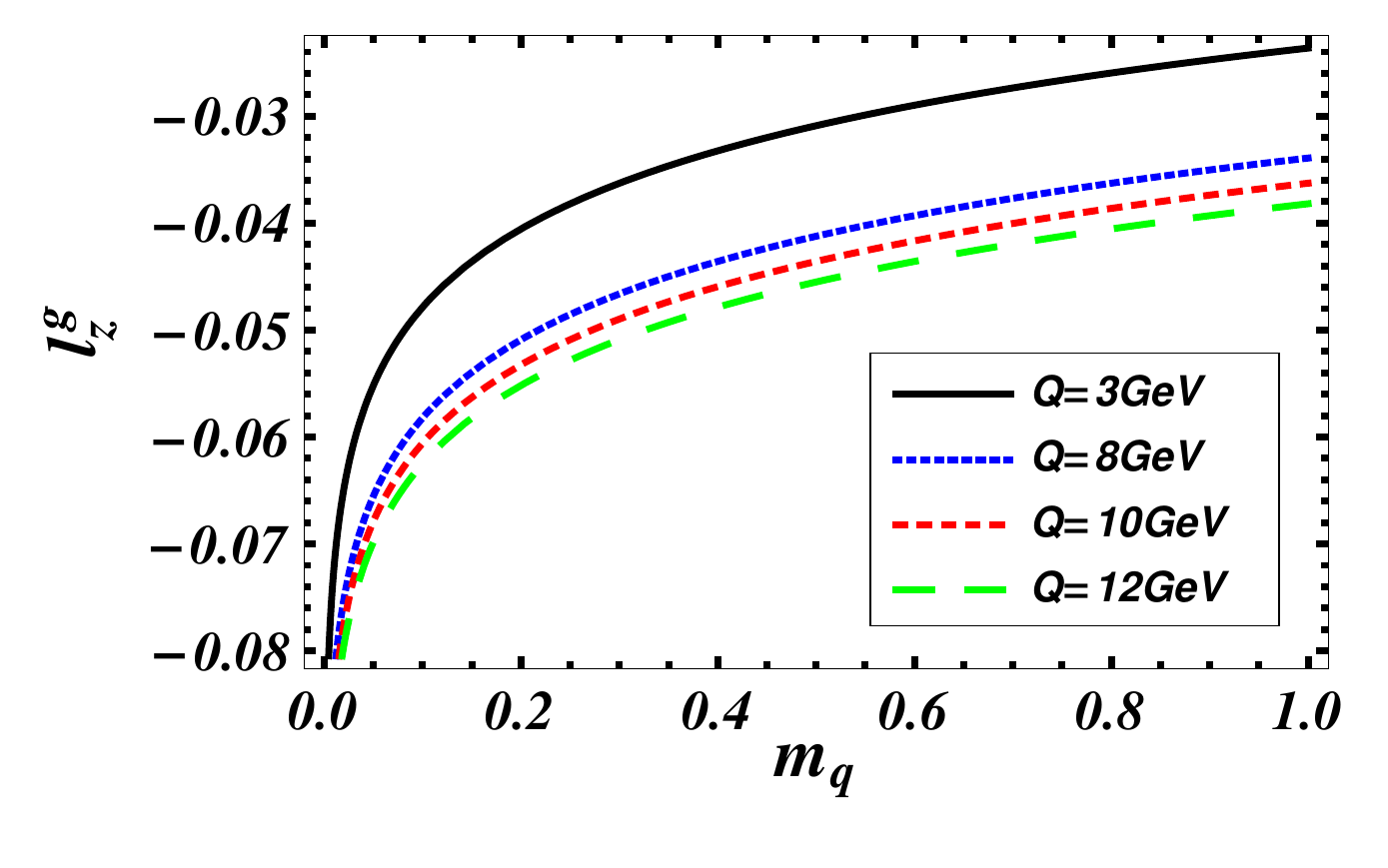}
\hspace{0.1cm}
\tiny{(b)}\includegraphics[width=7cm,height=6.5cm,clip]{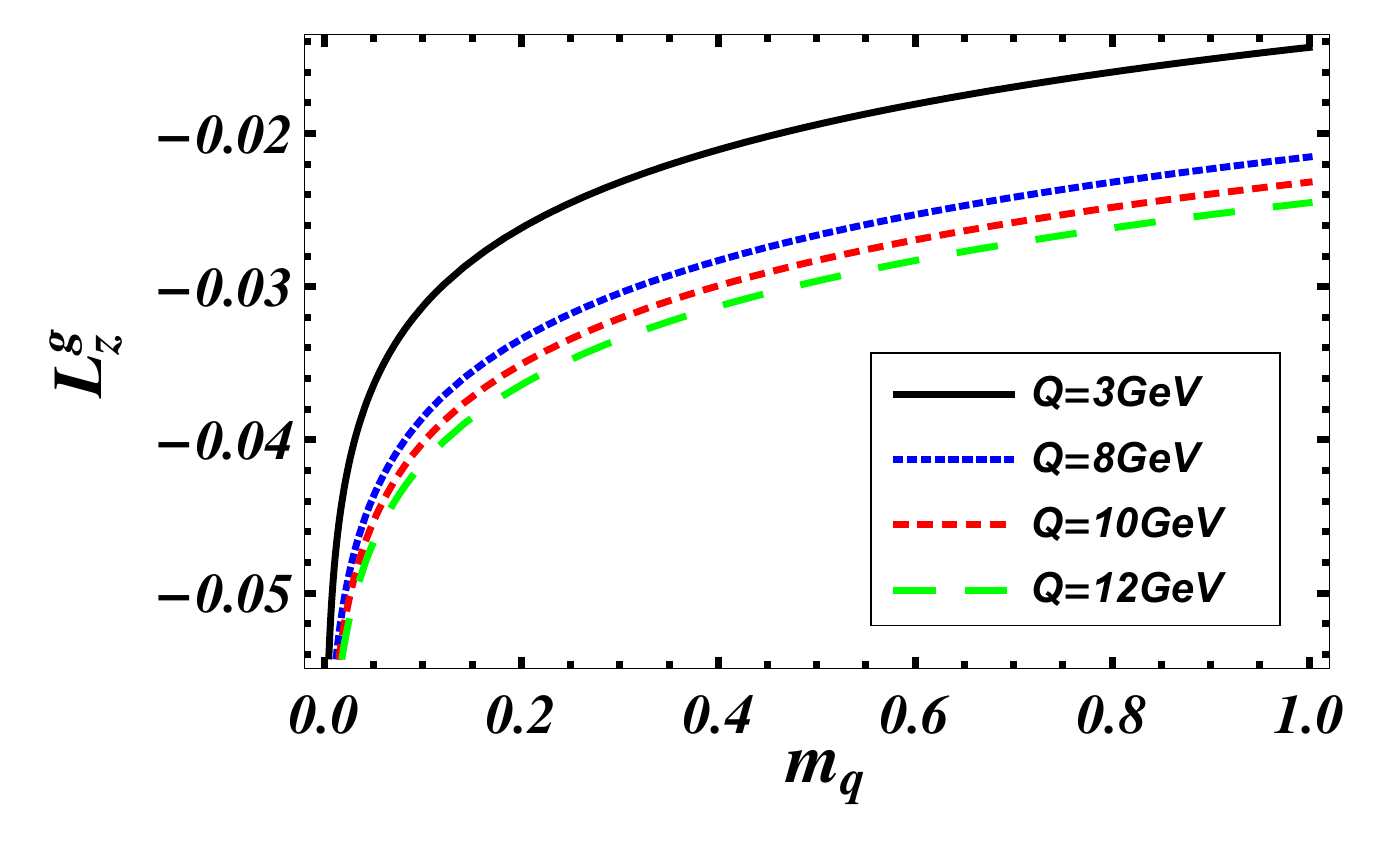}
\end{minipage}
\caption{\label{fig4}(Color online)
Plots of gluon OAM (a) canonical  and (b) kinetic  vs mass of the active
quark (GeV). $Q$ is the upper limit of the transverse momentum in GeV 
\cite{ours2}}
\end{figure}

In order to calculate the gluon GTMDs we use the parametrization of  in
\cite{gluon_gtmd}. These can be related to the ones in \cite{metz}.
Using this parametrization, one can relate the gluon GTMDs to the gluon
kinetic and canonical OAM in the same way as for the quarks. In our model,
the quark and gluon OAM  can be calculated directly from the results for the
Wigner distributions, as these are related to the GTMDs  \cite{ours1,ours2}. 
Our results agree with \cite{hari} in
the massless limit of the quark. We also agree with \cite{hikmat} and
\cite{other}. In our model
calculation we find that the GTMDs $F_{14}$ and $G_{11}$ exist and non-zero. The
result is independent of the choice of the gauge link at this order
\cite{other}. In Fig.\ref{fig3} we have shown the orbital angular momentum
of quarks as a function of the mass as calculated in this model.
 Fig. 3(a) is for the kinetic OAM and 3(b) for canonical OAM. 
$Q$ is the upper limit in the transverse momentum  integration, which 
is the large momentum scale involved in the process.
Similar qualitative behavior of $L^{q}_z$ and $l^{q}_z$ are seen, however,
the magnitude of the two OAM differs, this is due to the gluonic degrees of
freedom in the model. In Figs.  4(a) and 4(b) we show the canonical and the
kinetic gluon OAM respectively as a function of the quark mass. As in the
quark case, we see  that the magnitude of both the OAM decreases with increasing mass
of target state.

The correlation between the quark spin
and its OAM is given by \cite{lorce,lorce_14},

\begin{eqnarray}
C^{q}_{z} = \int dx d^{2}k_{\perp} \frac{k_{\perp}^2}{m^2} G_{11}.
\end{eqnarray}

In this model $F_{14}$=$-G_{11}$, the above correlation is the same as the
canonical OAM for the quarks.
The gluon spin-orbit correlations can be related to the gluon GTMDs in a way
similar to the above. Unlike for the quarks, canonical gluon OAM and spin-orbit
correlations are different in this model.
Another point to note is that the spin-orbit correlation for the quark in the dressed quark is
negative. This is opposite to what is observed in chiral quark-soliton model
and constituent quark model, namely here the quark spin is anti-aligned with
its OAM.

%%%%%%%%%%%%%%%%%%%%%%%%%%%%%%%%%%%%%%%%%%%%%%%%%%%
\section{Conclusion}

We presented a recent calculation of the Wigner distributions for quarks and
gluons. We took a simple composite spin-1/2 system which has a gluonic degree of
freedom, namely a quark dressed with a gluon. 
We calculated the Wigner distributions both for unpolarized and
longitudinally polarized target and quarks and gluons and investigated  the
correlations in transverse momentum and position space. We compared and
contrasted the results for the quark distributions with calculations  in light cone
constituent quark model and light-cone chiral quark soliton model. For the
gluon Wigner distributions, ours \cite{ours2} are the first results. We also
calculated the kinetic and canonical quark and gluon OAM and the spin-orbit
correlations. 

\section{Acknowledgement}

This work has been done in collaboration with Sreeraj Nair and Vikash K.
Ojha \cite{ours1,ours2}. We thank the organizers of the QCD Evolution Workshop, 2015 at
Jefferson lab for the invitation and hospitality.

\end{document}